\documentclass[]{spie}  

\usepackage{amsmath,amsfonts,amssymb}
\usepackage{graphicx}
\usepackage[colorlinks=true, allcolors=blue]{hyperref}

\title{Recovering simulated planet and disk signals using SCALES aperture masking}

\author[a]{Mackenzie R. Lach}
\author[a]{Steph Sallum}
\author[b]{Ravinder Banyal}
\author[c]{Natalie Batalha}
\author[d]{Geoff Blake}
\author[e]{Tim Brandt}
\author[f]{Zackery Briesemeister}
\author[a]{Aditi Desai}
\author[g]{Josh Eisner}
\author[h]{Wen-fai Fong}
\author[i]{Tom Greene}
\author[j]{Mitsuhiko Honda}
\author[c]{Isabel Kain}
\author[h]{Charlie Kilpatrick}
\author[d]{Katherine de Kleer}
\author[k]{Michael Liu}
\author[c]{Bruce Macintosh}
\author[a]{Raquel Martinez}
\author[d]{Dimitri Mawet}
\author[g]{Brittany Miles}
\author[l]{Caroline Morley}
\author[m]{Imke de Pater}
\author[o]{Diana Powell}
\author[p]{Patrick Sheehan}
\author[c]{Andrew Skemer}
\author[n]{Justin Spilker}
\author[q]{Deno Stelter}
\author[r]{Jordan Stone}
\author[b]{Arun Surya}
\author[b]{Sivarani Thirupathi}
\author[g]{Kevin Wagner}
\author[s]{Yifan Zhou}

\affil[a]{University of California, Irvine}
\affil[b]{Indian Institute of Astrophysics}
\affil[c]{University of California, Santa Cruz}
\affil[d]{Caltech}
\affil[e]{University of California, Santa Barbara}
\affil[f]{NASA Goddard Space Flight Center}
\affil[g]{University of Arizona}
\affil[h]{Northwestern University}
\affil[i]{NASA Ames Research Center}
\affil[j]{Okayama University of Science}
\affil[k]{University of Hawai'i IfA}
\affil[l]{University of Texas at Austin}
\affil[m]{University of California, Berkeley}
\affil[n]{Texas A\&M University}
\affil[o]{University of Chicago}
\affil[p]{National Radio Astronomy Observatory}
\affil[q]{University of California Observatories}
\affil[r]{Naval Research Laboratory}
\affil[s]{University of Virginia}

\authorinfo{Further author information: (Send correspondence to M.R.L.)\\M.R.L.: E-mail: mrlach@uci.edu\\  S.S.: E-mail: ssallum@uci.edu}

\begin{document} 
\maketitle

\begin{abstract}
The Slicer Combined with Array of Lenslets for Exoplanet Spectroscopy (SCALES) instrument is a lenslet-based integral field spectrograph that will operate at 2 to 5 microns, imaging and characterizing colder (and thus older) planets than current high-contrast instruments. Its spatial resolution for distant science targets and/or close-in disks and companions could be improved via interferometric techniques such as sparse aperture masking. We introduce a nascent Python package, \texttt{NRM-artist}, that we use to design several SCALES masks to be non-redundant and to have uniform coverage in Fourier space. We generate high-fidelity mock SCALES data using the \texttt{scalessim} package for SCALES' low spectral resolution modes across its 2 to 5 micron bandpass. We include realistic noise from astrophysical and instrument sources, including Keck adaptive optics and Poisson noise. We inject planet and disk signals into the mock datasets and subsequently recover them to test the performance of SCALES sparse aperture masking and to determine the sensitivity of various mask designs to different science signals.  
\end{abstract}


\section{INTRODUCTION}
\label{sec:intro}

High-contrast imaging is a technique that can directly image circumstellar structures such as protoplanetary disks \cite{Esposito_2020, Hung_2015} and planets \cite{Marois_2008, Haffert_2019}. Direct imaging can provide valuable information including atmospheric composition \cite{Macintosh_2015}, planet formation history (e.g., Macintosh et al.~2015\cite{Macintosh_2015}), disk morphology \cite{Andrews_2018}, and disk-planet dynamics (e.g., Wagner et al.~2023 \cite{Wagner_2023}). However, direct imaging is challenging in practice due to the high contrast of very dim planets at relatively small angular separations from their very luminous hosts \cite{Guyon_2013}. As a result, high-contrast imaging is better suited to imaging younger, thermally bright planets. Currently, this method is valuable for its ability to detect and characterize exoplanets whose orbital geometries are not amenable to the transit or radial velocity techniques - i.e., those with face-on orbits as seen from Earth and systems with very large angular separations. To date, the NASA Exoplanet Archive\footnote{NASA Exoplanet Archive: \url{https://exoplanetarchive.ipac.caltech.edu/}} lists only about 65 out of more than 5500 confirmed planets as having been discovered via direct imaging. The next generation of ground- and space-based observatories will continue to improve high-contrast imaging techniques. As detection limits and spectroscopic capabilities are improved, these facilities will directly detect and characterize a wider diversity of planets and disks, including those that are currently detectable via indirect techniques.

SCALES (Slicer Combined with Array of Lenslets for Exoplanet Spectroscopy) is an upcoming lenslet-based integral field spectrograph (IFS) that is slated to see first light at the W. M. Keck Observatory in 2025 \cite{Stelter_2020, Skemer_2022}. It will be the first facility-class, high-contrast IFS to operate from 2 to 5 microns, which will push the limits of exoplanet detection and characterization to longer-wavelength, colder temperature regimes, where there is lower contrast between star and planet luminosity (see e.g. Sallum et al. in these proceedings \cite{Sallum_2023b}). This will expand the population of exoplanetary systems that can be studied via direct imaging, and will enable new, spectrally-dispersed studies of planet formation. The design of SCALES improves upon its predecessor, ALES (Arizona Lenslets for Exoplanet Spectroscopy), which has demonstrated the science case for SCALES by producing the first spatially-resolved mid-infrared (3 - 4 $\mu$m) spectra of exoplanetary systems \cite{Skemer_2015} and by detecting a protoplanet interacting with its protoplanetary disk by driving the formation of spiral arms \cite{Wagner_2023}.

High-contrast imaging studies of exoplanets typically utilize either coronagraphs or unobstructed imaging techniques paired with differential imaging \cite{Nielsen_2019, Langlois_2021, Stone_2018}. Coronagraphs are optical components that block the light of a host star to allow imaging of high-contrast, close-in companions \cite{Foo_2005}. Imaging observations using coronagraphs are limited by the inner working angle, which is typically greater than 2$\lambda$/D (e.g., Nielsen et al.~2019\cite{Nielsen_2019}; Juanola-Parramon et al.~2022\cite{Juanola_2022}). Non-redundant aperture masking (NRM) is a technique that achieves moderate contrast at high angular resolution (5-8 magnitudes at $\lesssim\lambda/D$). It is capable of surpassing the diffraction limit and overcoming the angular resolution challenges associated with coronagraphy and traditional imaging techniques \cite{Tuthill_2000, Tuthill_2018, Sallum_2017, Sturmer_2012}. 

In an NRM setup, a small mask with some number of holes in the instrument pupil plane blocks the telescope aperture, turning the telescope into an interferometer. In a fully non-redundant mask, each baseline formed by a pair of holes is unique in its length and orientation, and thus contributes uniquely to the Fourier transform of the interferogram. 
The phase of each baseline is then a linear combination of phase due to systematics (i.e., atmospheric and instrumental) and that which is intrinsic to the science target.
Closure phases and squared visibilities are observables that can be obtained from masked observations, which can be used to reconstruct an image from, or fit models to, the data. The closure phase quantity is measured from each trio of holes in the mask, which form closing triangles. By summing the phases around a closing triangle, the first-order systematic phases drop out and the intrinsic phase remains \cite{Baldwin_1986}. Closure phases are highly sensitive to asymmetries and are thus powerful tools for close-in companion detections. Squared visibilities give the power associated with each baseline and are sensitive to the size of the science target on the sky.

In this work, we test the performance of several NRM designs with SCALES, making progress in our ultimate goal of designing a set of masks to be integrated into SCALES. We utilize the \texttt{scalessim} \cite{Breisemeister_2020} and \texttt{SAMpy} \cite{Sallum_2022} python packages to simulate SCALES images and to compute their closure phases and squared visibilities. Contrast curves are produced for each mask with a point source science target to aid in noise characterization, and to assess their relative performance and ability to detect companion signals. Star-planet signals are also injected into these frames, and an MCMC algorithm is run to find best-fit companion models. Additionally, a disk model is injected. For both object types, we reconstruct images from the observables using the \texttt{SQUEEZE} package \cite{Baron_2010}. We write and use a python package called \texttt{NRM-artist} to generate the NRM designs used in this work.

\section{Mask Designs}
\label{sec:procedures}

Designing fully non-redundant aperture mask designs is tedious to do by hand, but can be made easier with computational tools. Each baseline that can be created by pairing two holes must be unique in length and angle; this becomes increasingly difficult as more holes are added to a design, or as hole size is increased. We develop a python package called \texttt{NRM-artist}\footnote{\texttt{NRM-artist} available on Github: \url{https://github.com/kenzielach/NRM-artist}} that can produce random, non-redundant mask designs of a given hole size for the Keck aperture. We use \texttt{NRM-artist} to generate a variety of mask designs with 6, 7, and 9 holes 1m in diameter. While \texttt{NRM-artist} allows more freedom in hole location in its general implementation, in this study we require holes to be centered on a mirror segment, and we omit the innermost ring of segments to avoid hole overlap with the secondary obscuration. 

\begin{figure} [ht]
\begin{center}
\begin{tabular}{c} 
\includegraphics[height=5.25cm]{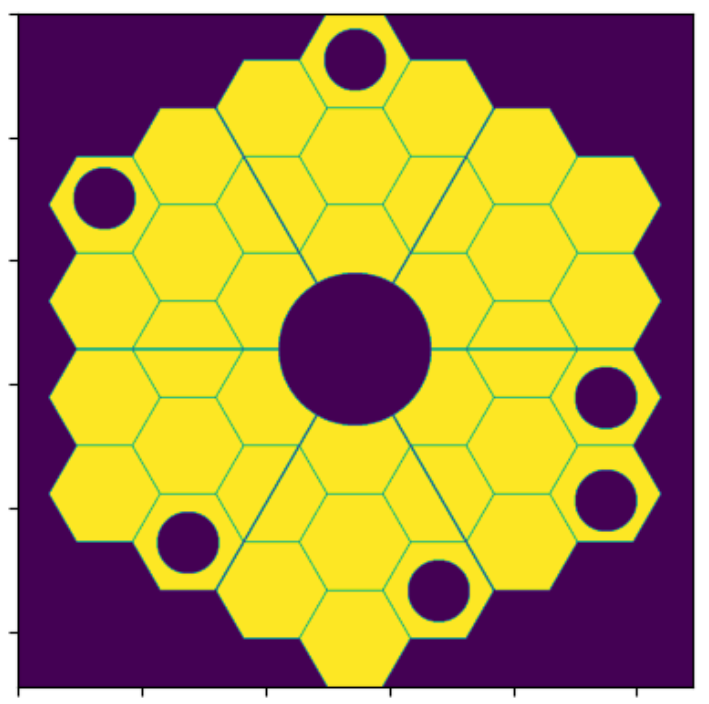}
\hfill
\includegraphics[height=5.25cm]{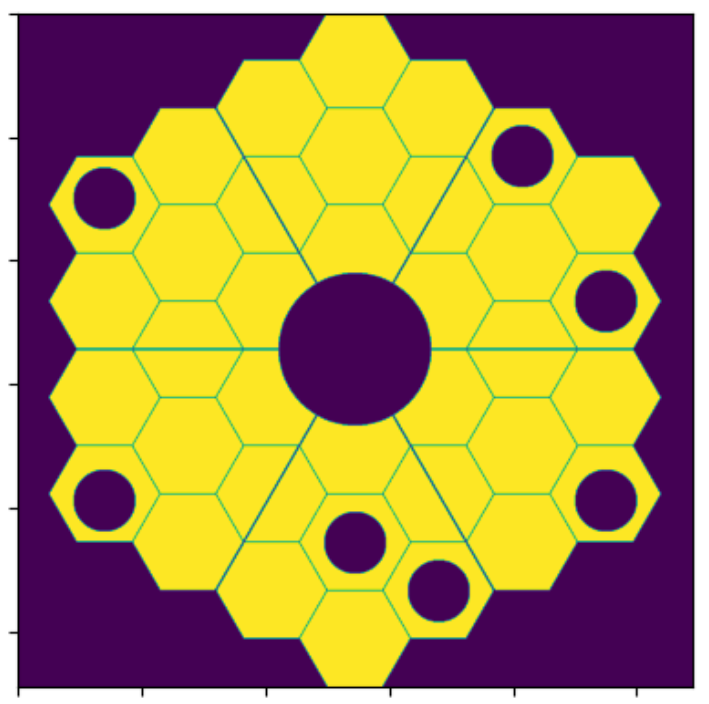}
\hfill
\includegraphics[height=5.25cm]{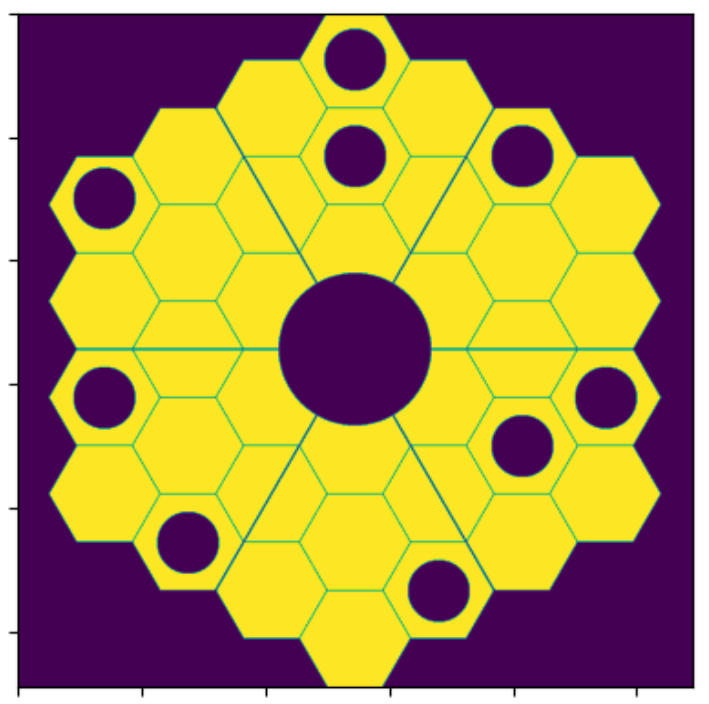}
\end{tabular}
\end{center}
\caption{Shown are the mask designs used in this work that were generated by \texttt{NRM-artist} and hand-selected based on broad and uniform u-v coverage. Holes are 0.5 m in radius and are depicted as projected onto the Keck primary.\label{fig:masks}}
\end{figure}

\texttt{NRM-artist} creates designs by randomly selecting hole positions as projected onto the Keck primary, and running checks to make sure no holes overlap with each other, with spiders, or with mirror segment gaps. It then runs a routine to check that the resulting mask design is completely non-redundant: proposed holes, with hole size and observing wavelength considered, are converted from x-y to u-v coordinates, and are checked to ensure there is no overlap. Currently, care must be taken to visually inspect mask designs in u-v space to make sure their Fourier coverage is uniform, and that baseline lengths take advantage of as much of the Keck primary diameter as possible. Three masks were selected to be used in this study based on these considerations: a 6-hole, a 7-hole, and a 9-hole. These mask designs are shown in Fig. \ref{fig:masks}. In future iterations of \texttt{NRM-artist}, functions will be implemented to automate the optimization of u-v coverage. Design generation is currently inefficient for masks with more than 7 holes under the aforementioned design requirements.

\section{Procedure for Simulating Observations}
\label{sec:sim_obs}
The \texttt{scalessim} package \footnote{\texttt{scalessim} available on Github: \url{https://github.com/scalessim/scalessim}; see Ref. ~\citenum{Breisemeister_2020} and Ref. ~\citenum{Sallum_2023b} for in-depth information.} was used to simulate a cube of SCALES low-resolution SED-mode (54 wavelength bins between 2.0 $\mu$m and 5.2 $\mu$m; R $\sim$ 35) images of a science scene for each mask design. Masked Keck PSF cubes were first generated at an oversampled wavelength resolution. Noise from Keck AO performance was simulated by applying an optical path difference (OPD) map to the PSF. These PSFs were next convolved with the input science scene. These were downsampled to match SCALES's low-resolution SED-mode. Using an integration time of 1 second, datacubes of simulated frames were generated with an FOV of $2.2^{\prime\prime}$ x $2.2^{\prime\prime}$ and with a plate scale of $0.02^{\prime\prime}$/spaxel.

\begin{figure} [ht]
\begin{center}
\begin{tabular}{c} 
\includegraphics[height=5.25cm]{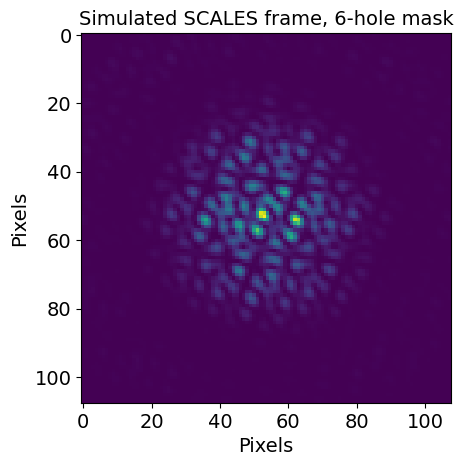}
\includegraphics[height=5.25cm]{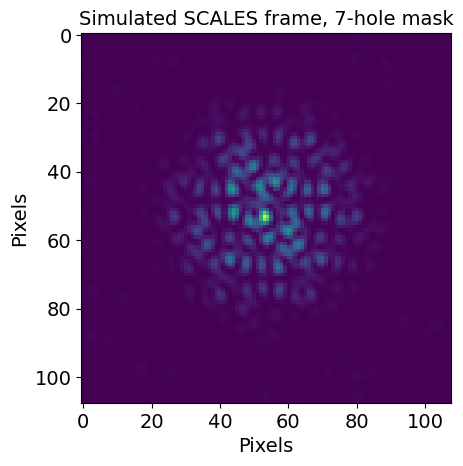}
\includegraphics[height=5.25cm]{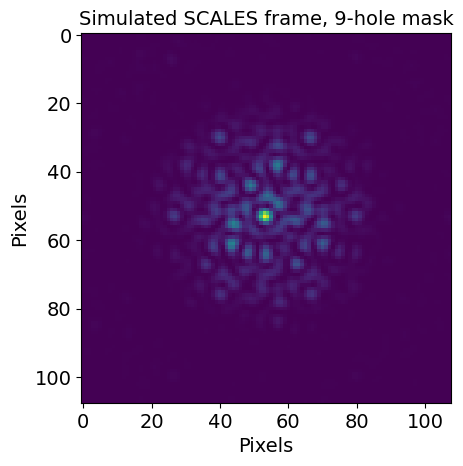}
\end{tabular}
\end{center}
\caption{Simulated images at L-band (3.509 $\mu$m) for each of the three mask designs tested. From left to right: 6-hole, 7-hole, 9-hole. The scene injected is of a star-planet companion model with a separation of 0.07$^{\prime\prime}$, stellar temperature of 2300 K, and planet temperature of 1500 K. \label{fig:rawims}
}
\end{figure}

Fig. \ref{fig:rawims} shows slices from simulated SCALES datacubes of an example injected planet model for each mask at L-band. The injected model separation was 0.07$^{\prime\prime}$, the stellar temperature was 2300 K , and the planet temperature was 1500 K . The system was modeled at a distance of 10 pc. Ten frames were taken per observation with random Poisson noise added to each to realistically simulate sky background noise. This process was repeated to generate frames for a reference star in order to simulate calibrator star observations. Calibration frames remove higher-order noise that is not eliminated by closure phases, but any mismatches in parameters such as target airmass, spectral type, and AO correction quality result in residual calibration errors \cite{Ireland_2013}. The calibration scene consists of the same star model as the target scene, with no companion. Poisson noise is added, and the PSF uses a version of the original OPD evolved forward in time to simulate quasi-static speckles. Background frames with Poisson noise but no star or companion were generated for both the target and calibrator scenes, using their corresponding OPD maps. After subtracting the background frames, closure phases and squared visibilities were calculated using the \texttt{SAMpy} package\footnote{\texttt{SAMpy} available on Github: \url{https://github.com/JWST-ERS1386-AMI/SAMpy}} and calibrated by subtracting the calibrator closure phases from the science target closure phases, and dividing the science target squared visibilities by the calibrator ones. 

To simulate reference PSF calibrators, the original OPD map used for the science target was evolved forward in time to simulate the changes in quasi-static speckles (the dominant source of calibration error in NRM observations\cite{Ireland_2013}) between a science target and a calibration star. Due to the frame rate of the AO system, it is computationally expensive to realistically simulate these maps over anything but very short exposure times. Ideally, we would perform an AO simulation for the full length of our integration time. For this study, we simplify the process of time evolution by artificially evolving a single OPD frame to obtain a single calibration OPD. We consider this a reasonable approximation since the error in NRM observations is dominated by calibration error. 

We begin with the original phase map calculated for the PyWFS \cite{Bond_2020} observing in the H-band with 0.66$^{\prime\prime}$ seeing. We expanded it using hexike polynomials, up to degree 30. The time evolution of the coefficients was simulated by randomly selecting new values within $\pm30\%$ of the original expanded coefficient values. The evolved OPD was then constructed from these values. In order to smooth out edge artifacts in the original, un-evolved OPD, it was expanded into its first 30 hexike terms, then reconstructed from those. This approach has been shown to introduce errors that are on the order of those expected in NIRC2 masking data \cite{Sallum_2019}. 

\begin{figure}[ht]
\begin{center}
\begin{tabular}{c} 
\\
\includegraphics[height=5.5cm]{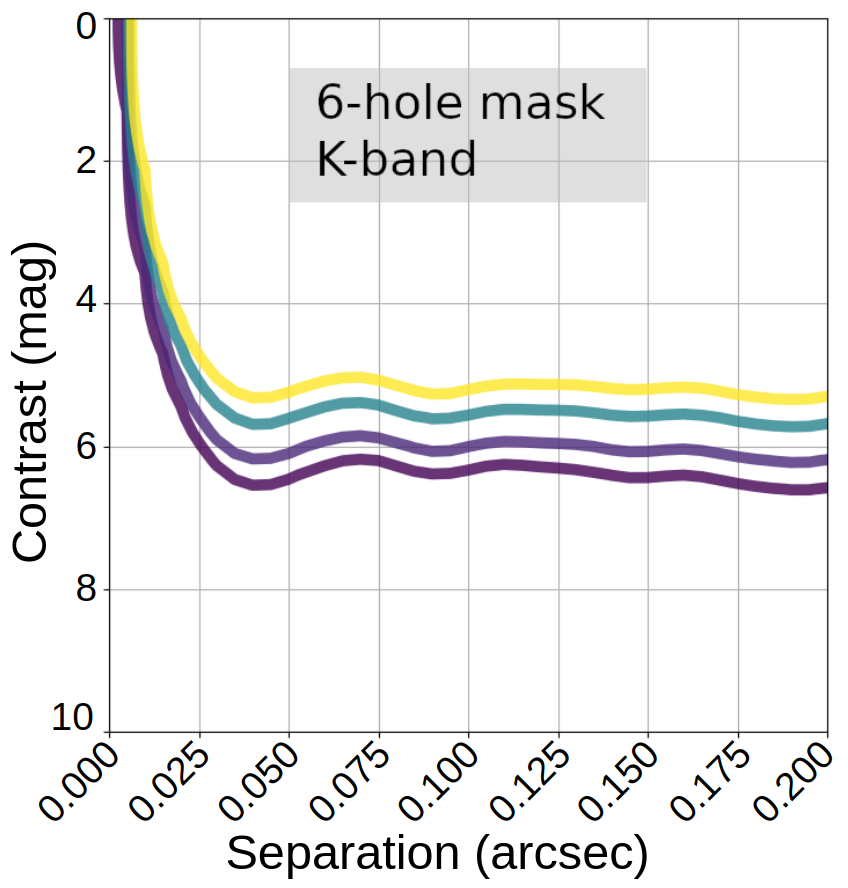}
\hfill
\includegraphics[height=5.5cm]{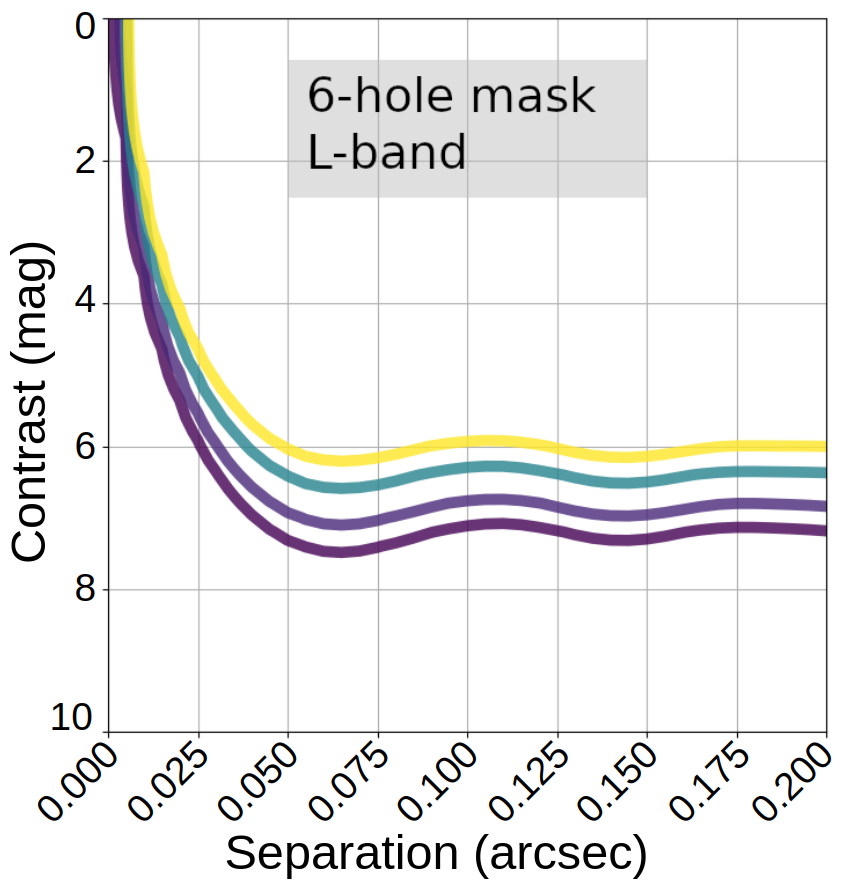}
\hfill
\includegraphics[height=5.5cm]{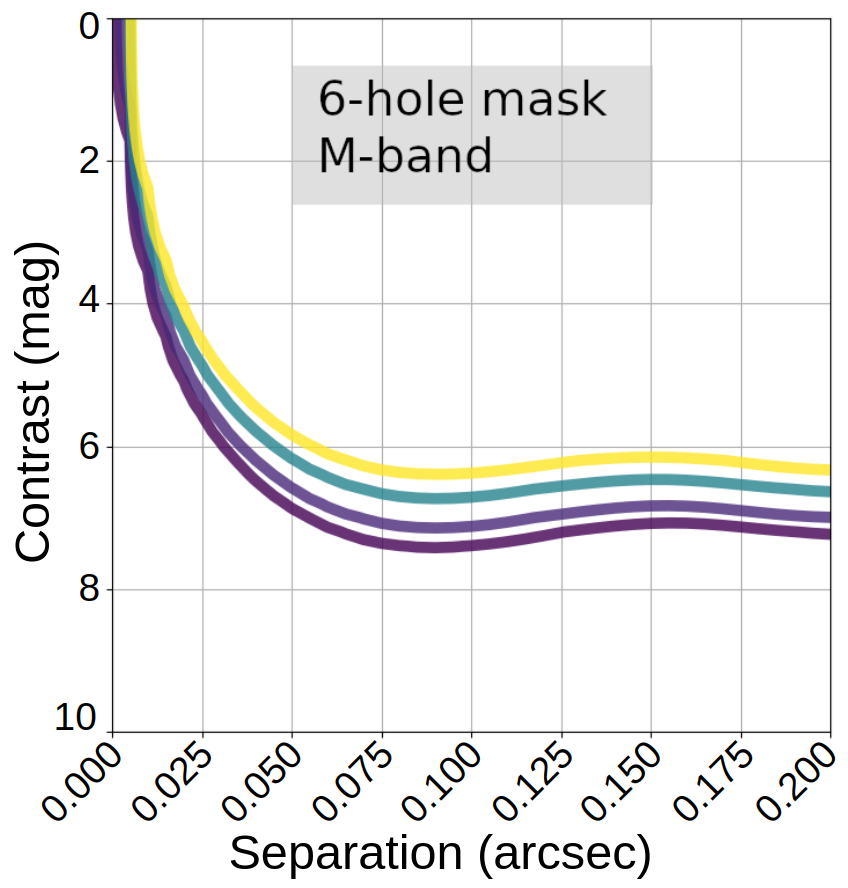}
\\
\includegraphics[height=5.5cm]{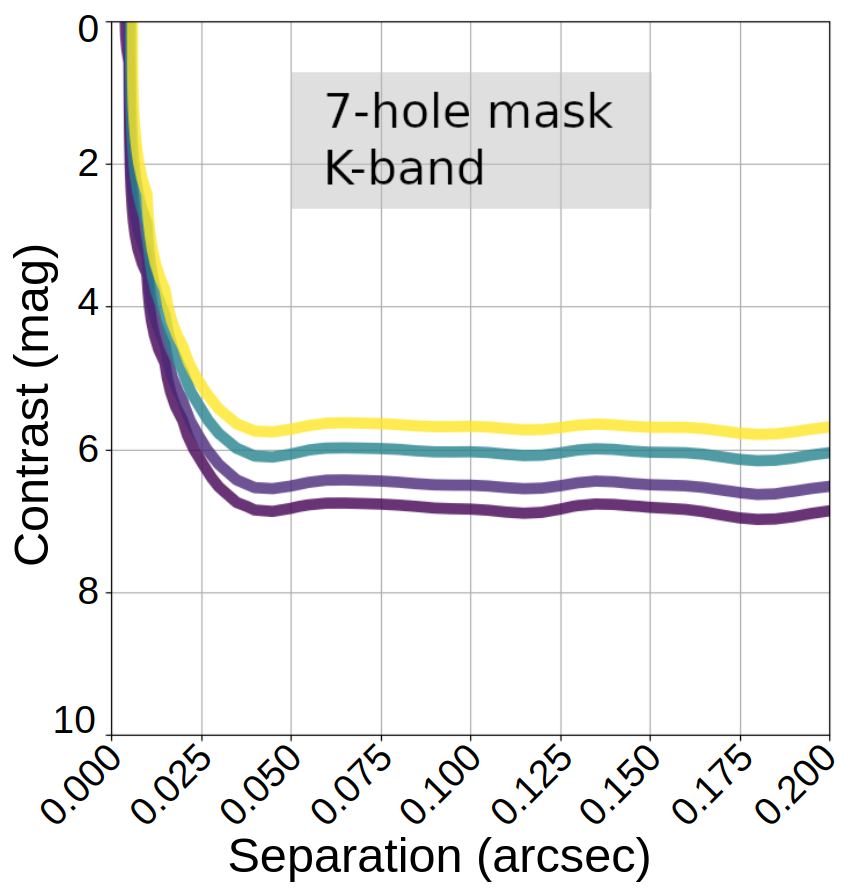}
\hfill
\includegraphics[height=5.5cm]{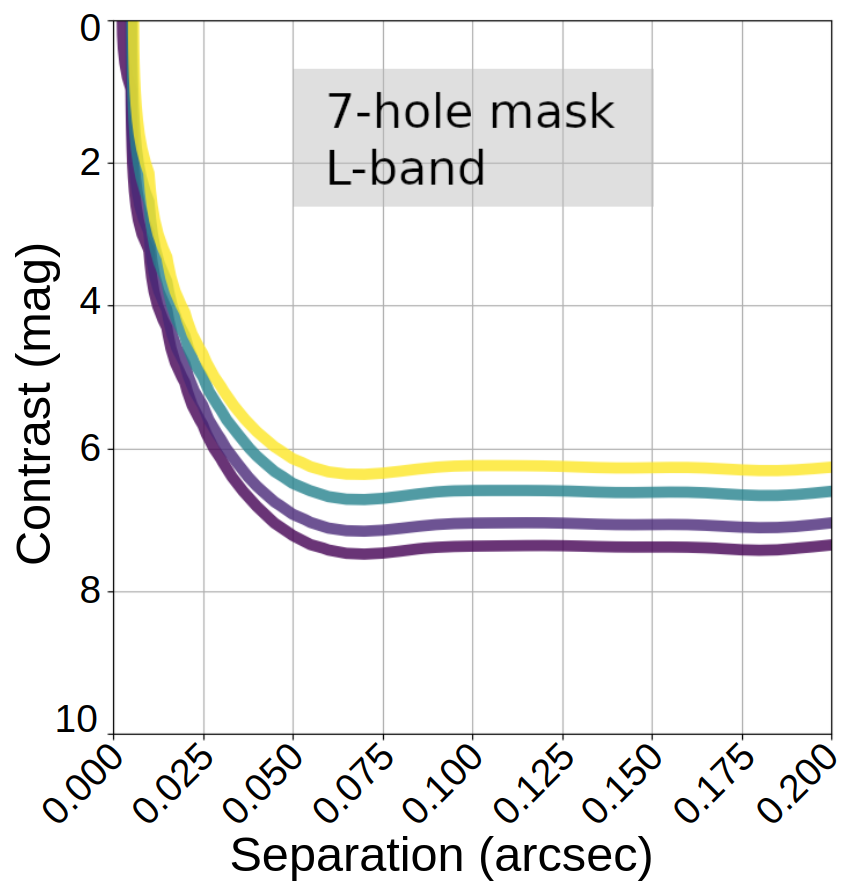}
\hfill
\includegraphics[height=5.5cm]{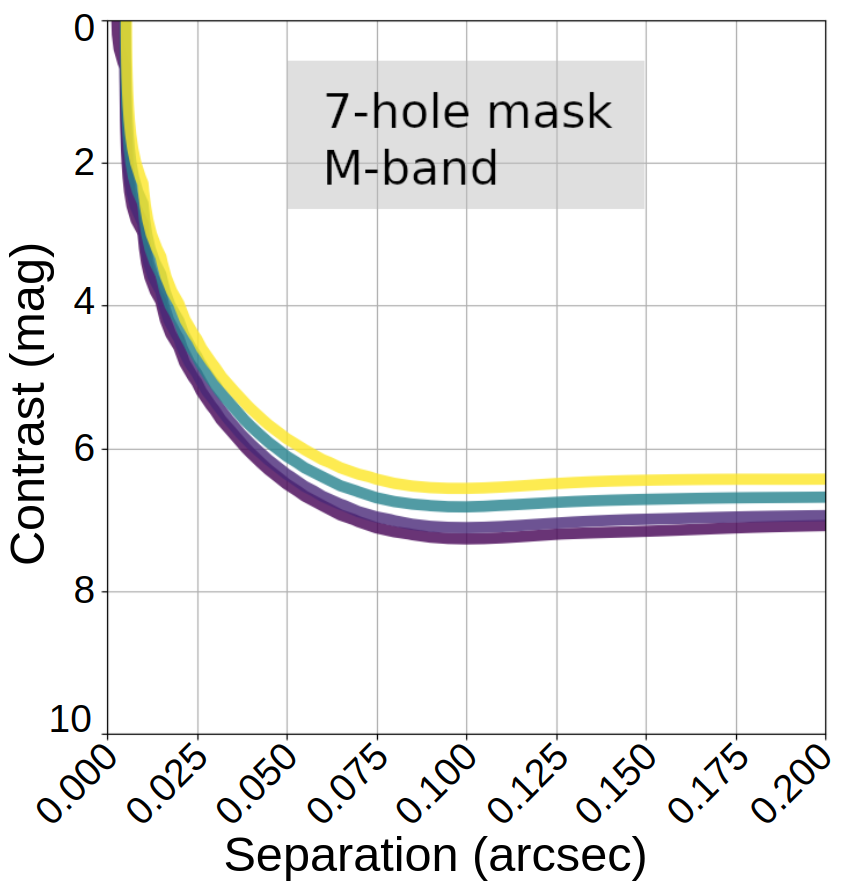}
\\
\includegraphics[height=5.5cm]{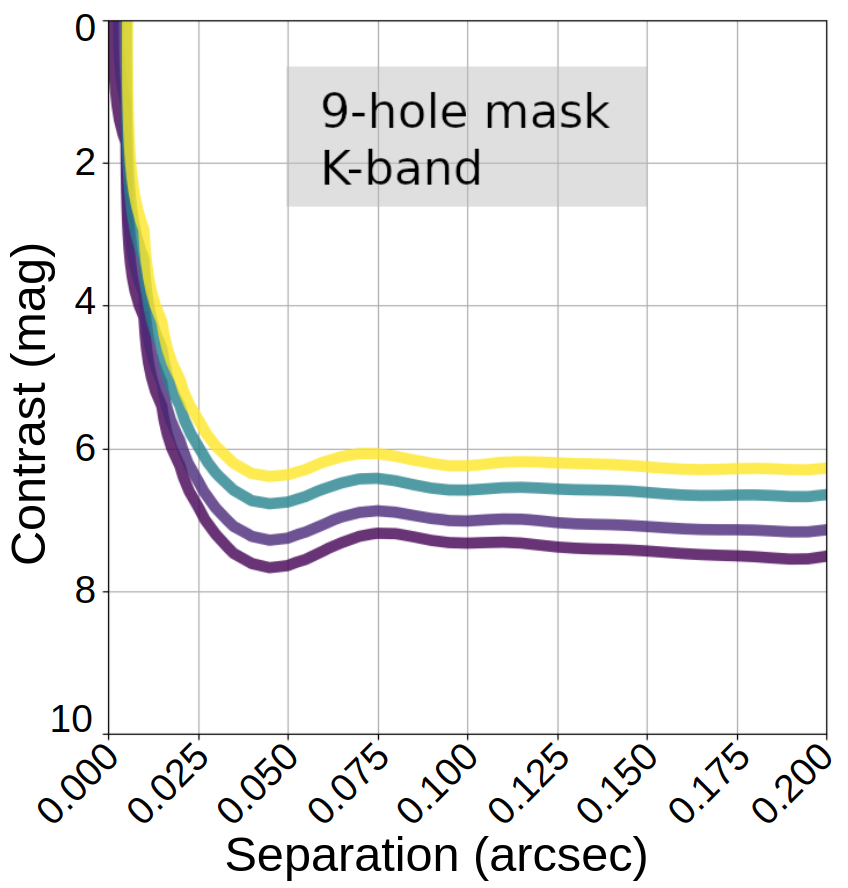}
\hfill
\includegraphics[height=5.5cm]{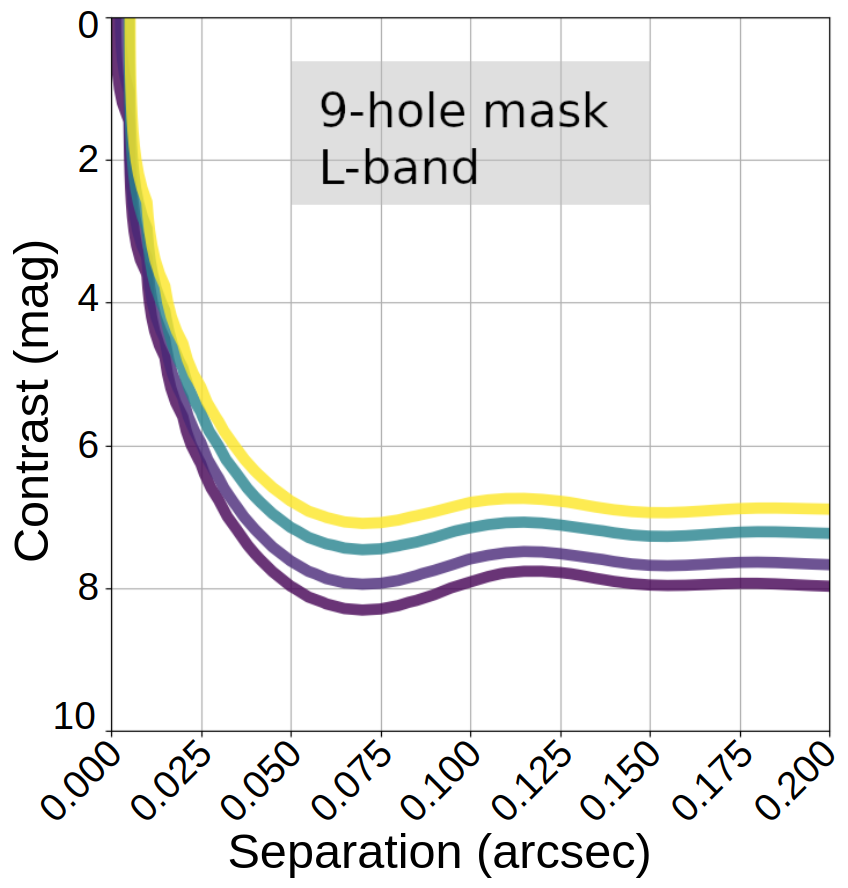}
\hfill
\includegraphics[height=5.5cm]{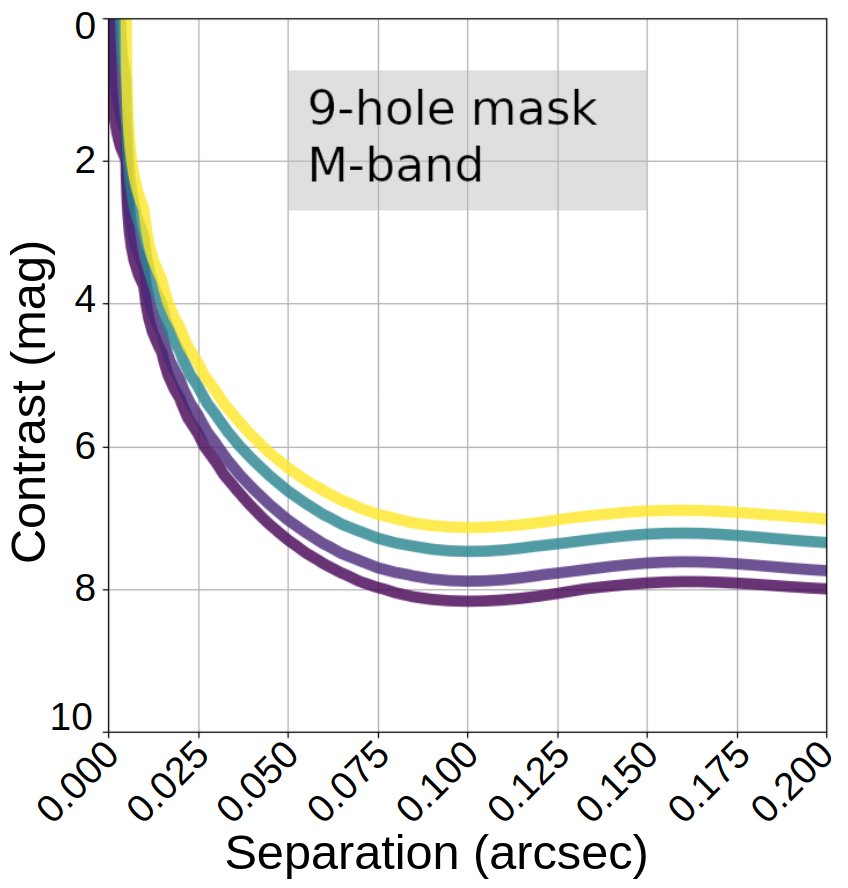}
\end{tabular}
\end{center}
\caption{Averaged contrast curves at sigma levels 1, 2, 5, and 10 (curves from bottom to top in each plot) for each of the three tested mask designs at K, L, and M bands. \label{fig:ccs}
}
\end{figure}

\begin{figure} [ht]
\begin{center}
\begin{tabular}{c} 
\includegraphics[height=5.4cm]{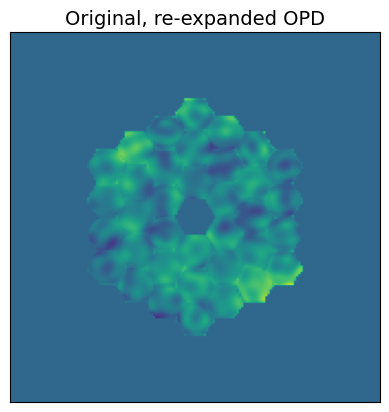}
\includegraphics[height=5.4cm]{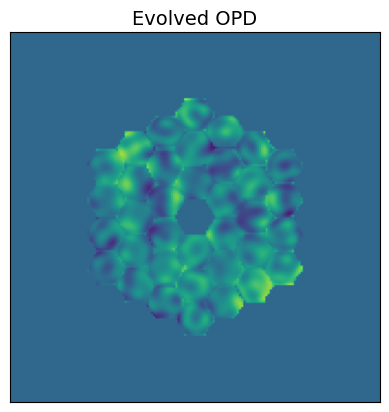}
\includegraphics[height=5.4cm]{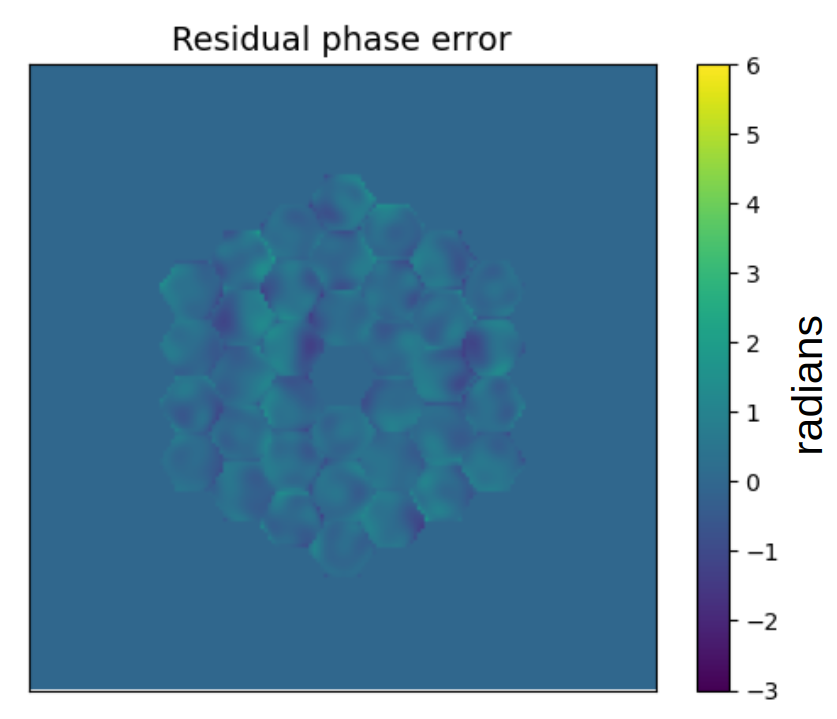}
\end{tabular}
\end{center}
\caption{Left: original, re-expanded (to 30$^\mathrm{th}$ degree hexike polynomial). Center: example of an evolved OPD used in this work. Right: difference between original and evolved OPDs. \label{fig:opds}
}
\end{figure}

Ten different evolved OPDs were produced to explore the average level of calibration error for this level of OPD evolution. To quantify the calibration error, closure phases, squared visibilities, and their errors were generated for a point source science target using the original unevolved OPD. These observables were calibrated using each of the ten evolved OPD maps. Each set of calibrated closure phases and visibilities were fit to a binary model, from which chi-squared grids were calculated. From these grids, ten sets of contrast curves were generated, and each set was averaged to obtain a single contrast plot at K-, L-, and M-bands for each mask design  (Fig. \ref{fig:ccs}). The OPD whose own contrast curves best resembled the averaged contours was selected to be used as the calibrator OPD for the injection and recovery tests. Fig. \ref{fig:opds} shows the original re-expanded OPD alongside the evolved map used to calibrate disk and planet injections in this study, as well as the residual phase error between the two maps.

A vital aspect of testing mask design performance is to attempt to recover injected signals and reconstruct a science image. In this work, one ``easy-to-detect" planet companion and one disk signal were injected into the simulated raw SCALES frames. We have only simulated a snapshot at one parallactic angle for these injection tests. We fit single companion models to the injected planet scene. We also reconstructed images for both scenes using the \texttt{SQUEEZE}\cite{Baron_2010} package, which takes in our calibrated closure phases and squared visibilities and runs an MCMC algorithm to find a best-fit reconstructed image.
The disk signal was generated using a geometric Gaussian ring model, which allows for disks of a given inner and outer radius, radial and azimuthal brightness profiles, and axis ratio (see Ref.~\citenum{Sallum_2023a} for details). A disk with major radius 100 mas, axis ratio of 0.674, and a fractional stellar flux of 0.75 was injected. 

For the planet signal, the science scene datacube was created with a central 2300 K Phoenix model \cite{Husser_2013} star and a 1500 K planet spectrum with a gravity of 316 kg m/s$^2$ from the Sonora Bobcat model \cite{Marley_2021}, at a separation of 0.07$^{\prime\prime}$. The target system is set to be at a distance of 10 pc. This separation is roughly equal to $\lambda$/D at L-band. SCALES frames were generated using this scene as described above with an integration time of 1 second, and calibrated closure phases and squared visibilities were calculated. These observables were passed to the \texttt{ptemcee}\cite{Vousden_2016} parallel-tempered Monte Carlo Markov Chain (MCMC) algorithm, which searched for a best-fit companion model with a given contrast, separation, and position angle. Parallel tempering is beneficial for situations such as this with multi-modal probability distributions, as it allows multiple chains of different temperatures to simultaneously jump between modes and explore individual modes, all while exchanging information with each other \cite{Vousden_2016}.

\section{Results}
\label{sec:results}

Fig. \ref{fig:ccs} shows the resulting contrast curve for each mask design and at individual wavelength bins in SCALES' K-, L-, and M-bands, at various $\sigma$ levels. As plotted, a deeper curve indicates a larger achievable contrast for given companion separation. Overall, these curves follow the expected trend of improving with increasing the number of mask holes. This is due to additional baselines sampling more points in Fourier space, and thus providing more information about the science scene. The curves tend to deepen at larger separations from K- to L-band, and worsen slightly from L- to M-band. The minor worsening at M-band could be caused by higher sky background. The performance loss at K-band could be due to a lower Strehl ratio at the shorter wavelength. However, at K-band slightly larger contrasts can be achieved for closer companions. This is because $\lambda$/D scales directly with wavelength, with companion separations on the order of or larger than this value being easier to detect.

\begin{figure}[ht]
\begin{center}
\begin{tabular}{c} 
\\
\includegraphics[height=4.2cm]{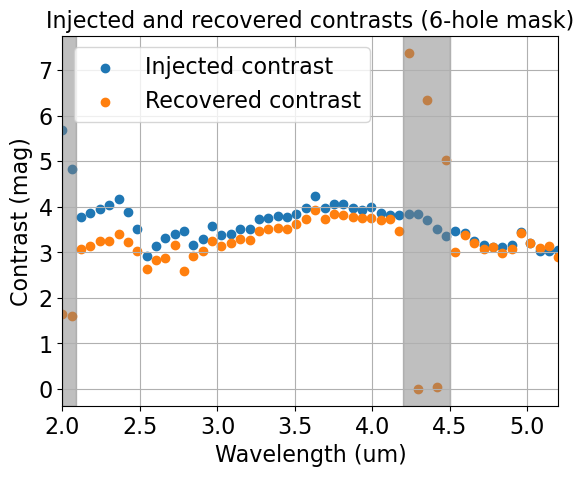}
\hfill
\includegraphics[height=4.2cm]{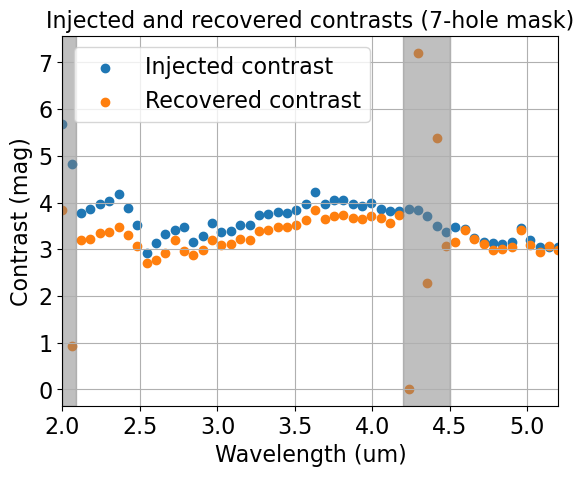}
\hfill
\includegraphics[height=4.2cm]{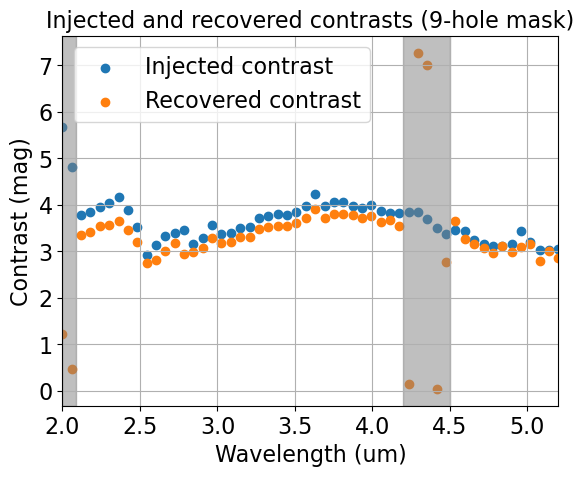}
\end{tabular}
\end{center}
\caption{Best-fit recovered contrasts versus wavelength for each mask design, plotted alongside the injected signal contrast. \label{fig:best_conts}
}
\end{figure}

\begin{figure}[ht]
\begin{center}
\begin{tabular}{c} 
\\
\includegraphics[height=5.5cm]{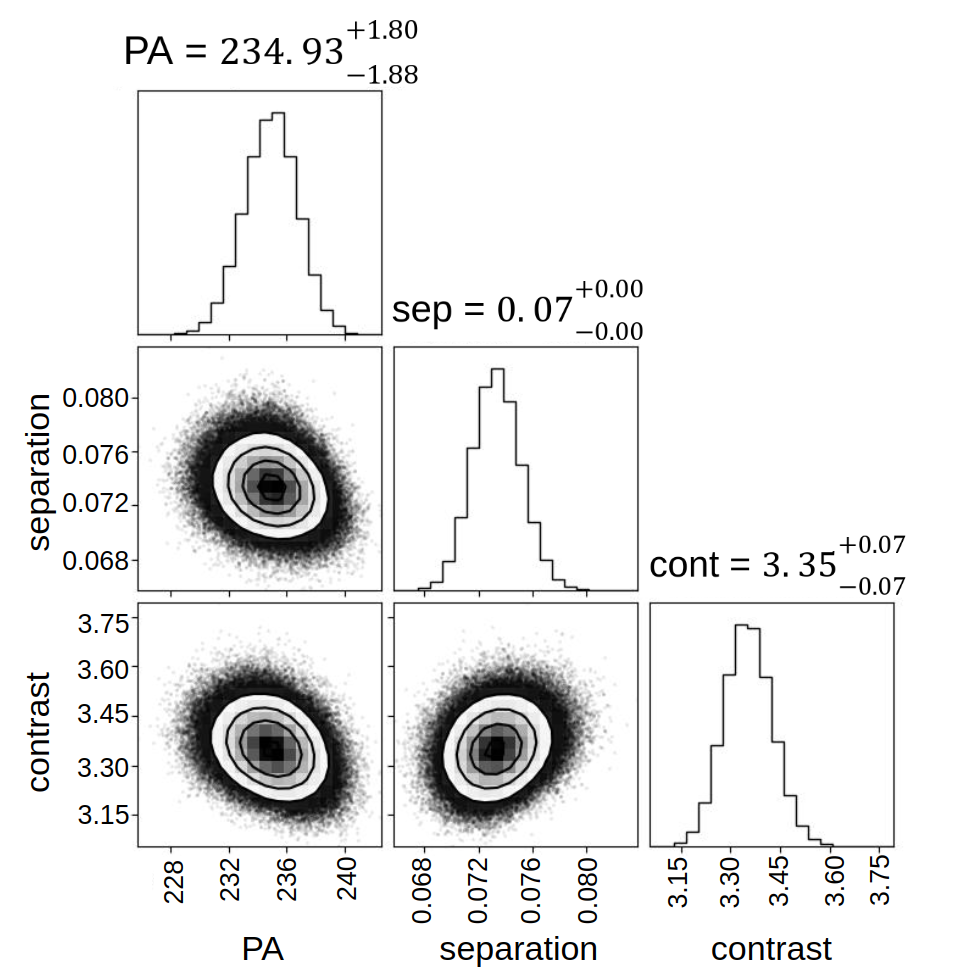}
\hfill
\includegraphics[height=5.5cm]{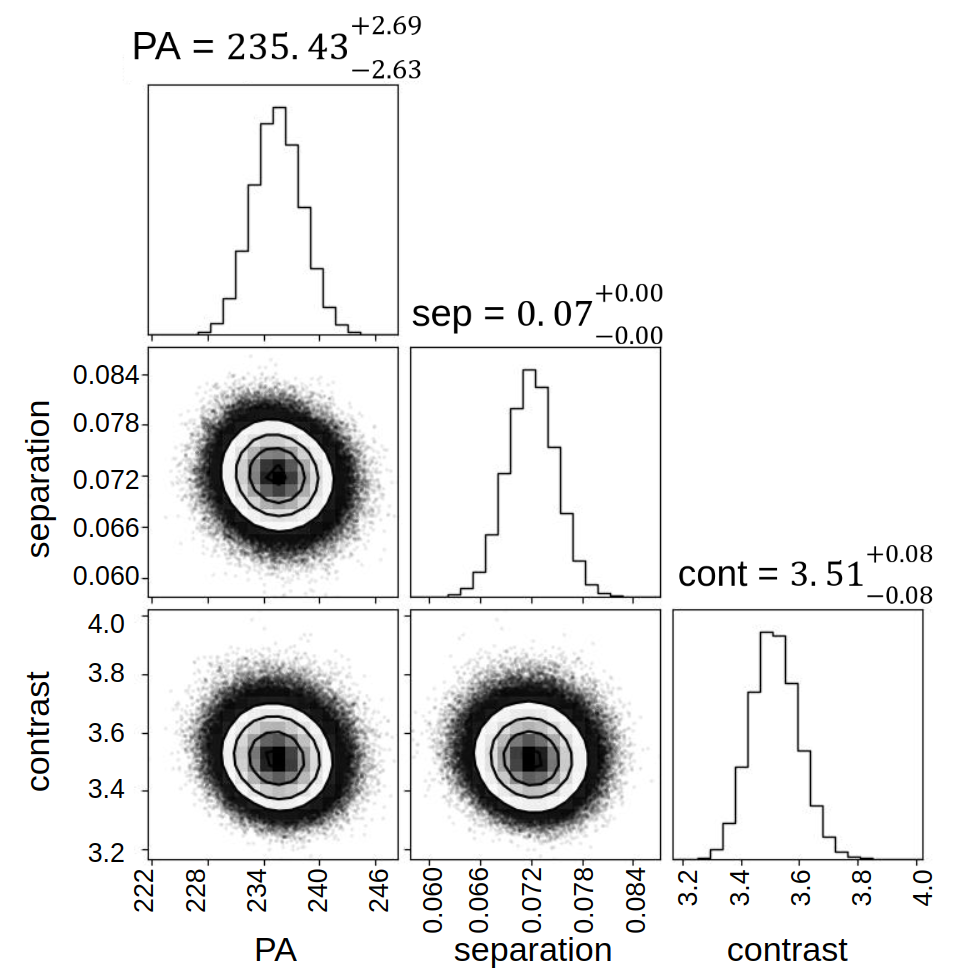}
\hfill
\includegraphics[height=5.5cm]{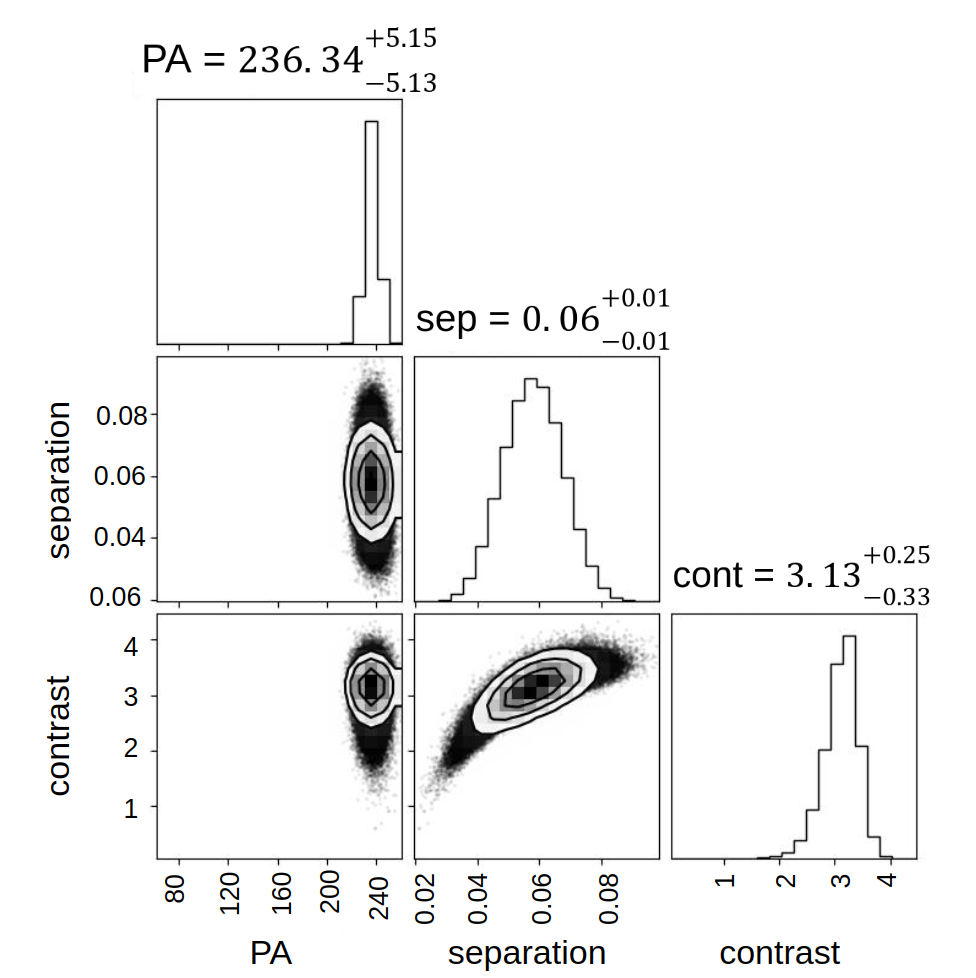}
\end{tabular}
\end{center}
\caption{Corner plots showing chi-squared surfaces and gaussians for best-fits of the 7-hole mask design at K-, L-, and M-bands. \label{fig:corners}
}
\end{figure}

Fig. \ref{fig:best_conts} shows both the injected planet signal contrast and the best-fit contrast, as retrieved by the MCMC algorithm, as a function of wavelength bin for each mask design. With the exception of two regions (shaded in Fig. \ref{fig:best_conts}), one at K-band where injected contrast increases and achievable contrast worsens, and the other a bin with very low atmospheric transmission, the fits generally recover the injected signals. K-band retrievals tend to be worse, perhaps due to the lower Strehl, although they can be seen to improve noticeably when more mask holes are added. Future work will address the inconsistency between injected and recovered contrasts at K-band, as it may lead to errors in photometric measurements up to a magnitude. Corner plots for three of these recovered fits using the 7-hole mask at K-, L-, and M-band wavelength bins are shown in Fig. \ref{fig:corners}. It is interesting to note the slight degeneracy in the M-band corner plot; this is due to the injected separation being inside of $\lambda$/D at that bin.

\begin{figure}[ht]
\begin{center}
\begin{tabular}{c} 
\\
\includegraphics[height=5.5cm]{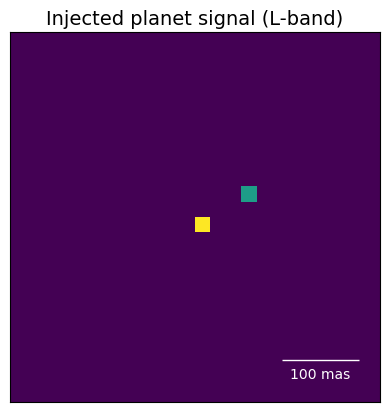}
\hfill
\includegraphics[height=5.5cm]{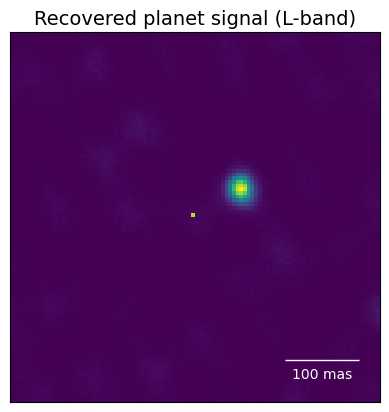}
\\
\includegraphics[height=5.5cm]{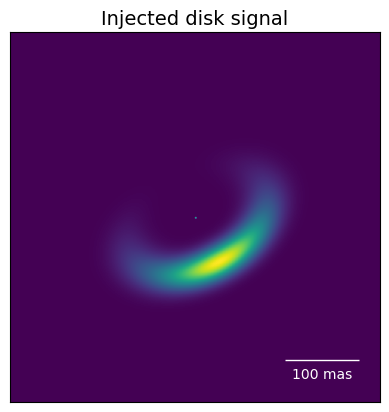}
\hfill
\includegraphics[height=5.5cm]{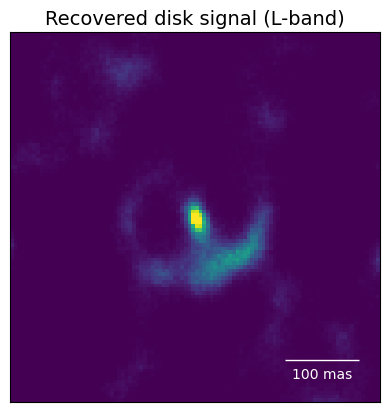}
\end{tabular}
\end{center}
\caption{The injected planet and disk signals (before PSF convolution) (left) and the reconstructed images recovered from simulated SCALES observations (right) are shown here. Images were recovered using the \texttt{SQUEEZE} software package \cite{Baron_2010}. Contrasts between star and companion are reduced in these images so the companion is easy to see. Note the planet and star in the injected planet signal (top left) are delta functions, which is why only single pixels are illuminated. \label{fig:images}
}
\end{figure}

Finally, the reconstructed images for our two science targets are shown alongside the injected scenes in Fig. \ref{fig:images}. These reconstructions closely resemble the geometry of the injected science scenes, including the asymmetry in the disk flux. However, some noise is clearly visible in the background of the reconstructed images. This is likely due to only one parallactic angle being taken into account in our simulations.

\section{Future Work}
\label{sec:future}
As described, the version of \texttt{NRM-artist} used in these Proceedings does not perform any optimization of mask designs. Future work will include further development of \texttt{NRM-artist} to add the functionality to efficiently search for mask designs with ideal u-v space coverage for SCALES. Additional features will be added to streamline and speed up its performance, to allow for a variety of mask design criteria, and to allow for greater flexibility in designing non-redundant aperture masks for telescopes other than Keck. An improved version of \texttt{NRM-artist} will be used to ultimately select several mask designs that will be integrated onto SCALES.

We intend to test these future mask designs with a greater variety of planet and disk signals, including colder planets around a wider range of stars. Joint fitting of models to each individual wavelength bin simultaneously will be done, as opposed to the less computationally expensive technique used in this study of fitting models to one bin at a time. Future calibrations will also take wavelength information into account (e.g., see Chaushev et al. ~2023\cite{Chaushev_2023}). Improvements will be made upon our techniques for generating SCALES PSFs, in order to simulate more realistic observations based in Keck AO performance predictions (including for planned upgrades). In future work, varying degrees of sky rotation will be taken into account. Mask designs for other SCALES low-resolution modes will also be tested.

\acknowledgments The authors would like to extend gratitude to Dr. Maaike van Kooten for providing the Keck OPD data utilized in this work. We are grateful to the Heising-Simons Foundation, the Alfred P. Sloan Foundation, and the Mt. Cuba Astronomical Foundation for their generous support of our efforts. This project also benefited from work conducted under the NSF Graduate Research Fellowship Program. S. S. is supported by the National Science Foundation under MRI Grant No. 2216481. Finally, we would like to acknowledge the Code Astro workshop at Northwestern University for providing M. R. L. with time, guidance, and opportunity to begin developing NRM-artist.


\bibliography{report} 
\bibliographystyle{spiebib} 

\end{document}